\documentclass[conference,a4paper]{IEEEtran}
\usepackage[utf8]{inputenc} 
\usepackage[T1]{fontenc}
\usepackage{url}
\usepackage{ifthen}
\usepackage{cite}
\usepackage[cmex10]{amsmath} % Use the [cmex10] option to ensure complicance
\usepackage{amsfonts,amssymb}
\usepackage[font=scriptsize,caption=false,labelsep=space]{subfig}
\usepackage{graphicx,float,wrapfig}
\usepackage{graphicx}
\usepackage{graphics,color}
\usepackage{mathtools}
\usepackage{nicefrac}
\usepackage{bm}
\usepackage{url}

\usepackage{balance}

\usepackage[noend]{algpseudocode}
\usepackage{algorithmicx}
\usepackage{algorithm}
\usepackage{color}

\usepackage{wrapfig} % Allows in-line images

\usepackage{upquote}
\usepackage[none]{hyphenat}

\newtheorem{thm}{Theorem}
\newtheorem{theorem}[thm]{Theorem}

\newtheorem{corollary}[thm]{Corollary}

\begin{document}
\title{Information Geometric Approach to Bayesian Lower Error Bounds} 

%%% Several authors with up to three affiliations:
\author{%
  \IEEEauthorblockN{M.~Ashok Kumar}
  \IEEEauthorblockA{Discipline of Mathematics\\
                    Indian Institute of Technology\\
                    Indore, Madhya Pradesh, India\\
                    Email: ashokm@iiti.ac.in}
  \and
  \IEEEauthorblockN{Kumar Vijay~Mishra}
  \IEEEauthorblockA{IIHR - Hydroscience \& Engineering\\
                    The University of Iowa\\ 
                    Iowa City, IA, USA\\
                    Email: kumarvijay-mishra@uiowa.edu}
}

\maketitle

\begin{abstract}
Information geometry describes a framework where probability densities can be viewed as differential geometry structures. This approach has shown that the geometry in the space of probability distributions that are parameterized by their covariance matrix is linked to the fundamentals concepts of estimation theory. In particular, prior work proposes a Riemannian metric - the distance between the parameterized probability distributions - that is equivalent to the Fisher Information Matrix, and helpful in obtaining the deterministic Cram\'{e}r-Rao lower bound (CRLB). Recent work in this framework has led to establishing links with several practical applications. However, classical CRLB is useful only for unbiased estimators and inaccurately predicts the mean square error in low signal-to-noise (SNR) scenarios. In this paper, we propose a general Riemannian metric that, at once, is used to obtain both Bayesian CRLB and deterministic CRLB along with their vector parameter extensions. We also extend our results to the Barankin bound, thereby enhancing their applicability to low SNR situations.
\end{abstract}

\section{Introduction}
\label{sec:intro}
Information geometry is a study of statistical models from a Riemannian geometric perspective. The differential geometric modeling methods were introduced to statistics by C. R. Rao in his seminal paper \cite{rao1945information} and later formally developed by Cencov \cite{cencov2000statistical}. The information geometric concept of a manifold that describes the parameterized probability distributions has garnered considerable interest in recent years. The main advantages include structures in the space of probability distributions that are invariant to non-singular transformation of parameters \cite{amari2000methods}, robust estimation of covariance matrices \cite{balaji2014information} and usage of Fisher Information Matrix (FIM) as a metric \cite{nielsen2013cramer}. 

Information geometry has now transcended its initial statistical scope and expanded to several novel research areas including, but not limited to, Fisher-Rao Riemannian geometry \cite{maybank2012fisher}, Finsler information geometry \cite{shen2006riemann}, optimal transport geometry \cite{gangbo1996geometry}, and quantum information geometry \cite{grasselli2001uniqueness}. Many problems in science and engineering use probability distributions and, therefore, information geometry has been used as a useful and rigorous tool for analyses in applications such as neural networks \cite{amari1997information,amari2002information}, optimization \cite{amari2013minkovskian,beyer2014convergence}, radar systems \cite{de2014design,barbaresco2008innovative}, communications \cite{coutino2016direction}, computer vision \cite{maybank2012fisher}, and machine learning \cite{sun2014information,amari1998natural}. More recently, several developments in deep learning \cite{desjardins2015natural,roux2008topmoumoute} that employ various approximations to the FIM to calculate the gradient descent have incorporated information geometric concepts.

The information geometry bases the distance between the parameterized probability distributions on the FIM \cite{amari2000methods}. In estimation theory, the well-known deterministic Cram\'{e}r-Rao lower bound (CRLB) is the inverse of FIM. Therefore, the results derived from information geometry are directly connected with the fundamentals of estimation theory. Nearly all prior works exploited this information geometric link to CRLB in their analyses because the CRLB is most widely used benchmark for evaluating the mean square error (MSE) performance of an estimator. However, the classical CRLB holds only if the estimator is unbiased. In general, the estimators are biased in many practical problems such as nonparametric regression \cite{gill1995applications}, communication \cite{prasad2013cramer}, and radar \cite{mishra2017performance}. The above-mentioned information geometric framework ceases its utility in these cases.

Moreover, the classical CRLB is a tight bound only when the errors are small. It is well known that in case of the nonlinear estimation problems with finite support parameters, for example the time delay estimation in radar \cite{mishra2017performance}, the performance of the estimator is characterized by the presence of three distinct signal-to-noise-ratio (SNR) regions \cite{van2013detection}. When the observations are large or the SNR is high (\textit{asymptotic} region), the CRLB describes the MSE accurately. In case of few observations or low SNR regions, the information from signal observations is insufficient and the estimator criterion
is hugely corrupted by the noise. Here, the MSE is close to that obtained via only \textit{a priori} information, that is, a quasi-uniform random variable on the parameter support. In between these two limiting cases lies the \textit{threshold} region where the signal observations are subjected to ambiguities and the estimator MSE increases sharply due to the outlier effect. The CRLB is used only in the asymptotic area and is not an accurate predictor of MSE when the performance breaks down due to increase in noise.

In this paper, we propose the information geometric framework that addresses these drawbacks of the classical CRLB. The Bayesian CRLB \cite{gill1995applications} is typically used for assessing the quality of biased estimators. It is similar to the deterministic CRLB except that it assumes the parameters to be random with an \textit{a priori} probability density function. We develop a general Riemannian metric that can be modified to link to both Bayesian and deterministic CRLB. To address the problem of the threshold effect, other bounds that are tighter than the CRLB have been developed (see e.g. \cite{renaux2008fresh} for an overview) to accurately identify the SNR thresholds that define the ambiguity region. In particular, Barankin bound \cite{barankin1949locally} is a fundamental statistical tool to understand the threshold effect for unbiased estimators. In simple terms, the threshold effect can be understood as the region where Barankin deviates from the CRLB \cite{knockaert1997barankin}. In this paper, we show that our metric is also applicable for the Barankin bound. Hence, compared to previous works \cite{amari2000methods}, our information geometric approach to minimum bounds on MSE holds good for both the Bayesian CRLB and deterministic CRLB, their vector equivalents and the threshold effect through the Barankin bound.

The paper is organized as follows: In the next section, we provide a brief background to the information geometry and describe the notation used in the later sections. Further, we explain the dual structure of the manifolds because, in most applications, the underlying manifolds are dually flat. Here, we define a divergence function between two points in a manifold. In Section \ref{sec:fim}, we establish the connection between the Riemannian metric and the Kullback-Leibler divergence for the Bayesian case. The manifold of all discrete probability distributions is dually flat and the Kullback-Leibler divergence plays a key role here. We also show that, under certain conditions, our approach yields the previous results from \cite{amari2000methods}. In Section \ref{sec:error_bounds}, we state and prove our main result applicable to several other bounds before providing concluding remarks in Section \ref{sec:conc}.

\section{Information Geometry: A Brief Background}
\label{sec:inf}
A $n$-dimensional manifold is a Hausdorff and second countable topological space which is locally homeomorphic to Euclidean space of dimension $n$ \cite{gallot2004riemannian,jost2005riemannian,spivak2005comprehensive}. A Riemannian manifold is a real differentiable manifold in which the tangent space at each point is a finite dimensional Hilbert space and, therefore, equipped with an inner product. The collection of all these inner products is called a Riemannian metric. 

In the information geometry framework, the statistical models play the role of a manifold and the Fisher information matrix and its various generalizations play the role of a Riemannian metric. Formally, by a {\em statistical manifold}, we mean a parametric family of probability distributions $S = \{p_\theta: \theta\in\Theta\}$ with a ``continuously varying" parameter space $\Theta$ (statistical model). The dimension of a statistical manifold is the dimension of the parameter space. For example, $S = \{N(\mu,\sigma^2) : \mu\in \mathbb{R}, \sigma^2 > 0\}$ is a two dimensional statistical manifold. The tangent space at a point of $S$ is a linear space that corresponds to a ``local linearization'' at that point. The tangent space at a point $p$ of $S$ is denoted by $T_p(S)$. The elements of $T_p(S)$ are called {\em tangent vectors} of $S$ at $p$. A {\em Riemannian metric} at a point $p$ of $S$ is an inner product defined for any pair of tangent vectors of $S$ at $p$.

In this paper, let us restrict to statistical manifolds defined on a finite set $\mathcal{X} = \{a_1,\dots,a_d\}$.  Let $\mathcal{P} := \mathcal{P}(\mathcal{X})$ denote the space of all probability distributions on $\mathcal{X}$. Let $S\subset\mathcal{P}$ be a sub-manifold. Let $\theta = (\theta_1,\dots,\theta_k)$ be a parameterization of $S$. Given a divergence function \footnote{By a {\em divergence}, we mean a non-negative function $D$ defined on $S\times S$ such that $D(p,q) = 0$ iff $p=q$.} on $S$, Eguchi \cite{eguchi1992geometry} defines a Riemannian metric on $S$ by the matrix
\[
G^{(D)}(\theta) = \left[g_{i,j}^{(D)}(\theta)\right],
\]
where
\begin{eqnarray*}
	g_{i,j}^{(D)}(\theta) & := & -D[\partial_i,\partial_j]\\
	& := & -\frac{\partial}{\partial\theta_i}\frac{\partial}{\partial\theta_j'}D(p_{\theta},p_{\theta'})\bigg|_{\theta=\theta'}
\end{eqnarray*}
where $g_{i,j}$ is the elements in the $i$th row and $j$th column of the matrix $G$, $\theta = (\theta_1,\dots,\theta_n)$, $\theta' = (\theta_1',\dots,\theta_n')$, and dual affine connections $\nabla^{(D)}$ and $\nabla^{(D^*)}$, with connection coefficients described by following Christoffel symbols
\begin{eqnarray*}
	\Gamma_{ij,k}^{(D)}(\theta) & := & -D[\partial_i\partial_j,\partial_k]\\
	& := & -\frac{\partial}{\partial\theta_i}\frac{\partial}{\partial\theta_j}\frac{\partial}{\partial\theta_k'}D(p_{\theta},p_{\theta'})\bigg|_{\theta=\theta'}
\end{eqnarray*}
and
\begin{eqnarray*}
	\Gamma_{ij,k}^{(D^*)}(\theta) & := & -D[\partial_k,\partial_i\partial_j]\\
	& := & -\frac{\partial}{\partial\theta_k}\frac{\partial}{\partial\theta_i'}\frac{\partial}{\partial\theta_j'}D(p_{\theta},p_{\theta'})\bigg|_{\theta=\theta'},
\end{eqnarray*}
such that, $\nabla^{(D)}$ and $\nabla^{(D^*)}$ form a dualistic structure in the sense that
\begin{eqnarray}
	\label{dualistic-structure}
	\partial_k g_{i,j}^{(D)}=\Gamma_{ki,j}^{(D)}+ \Gamma_{kj,i}^{(D^*)},
\end{eqnarray}
where $D^*(p,q) = D(q,p)$.

\section{Fisher Information Matrix for the Bayesian Case}
\label{sec:fim}
Eguchi's theory in section \ref{sec:inf} can also be extended to the space $\tilde{\mathcal{P}}(\mathcal{X})$ of all measures on $\mathcal{X}$. That is, $\tilde{\mathcal{P}} = \{\tilde{p}:\mathcal{X}\to (0,\infty)\}$. Let $S = \{p_\theta: \theta = (\theta_1,\dots,\theta_k)\in\Theta\}$ be a $k$-dimensional sub-manifold of $\mathcal{P}$ and let
\begin{eqnarray}
\label{eqn:denormalized_manifold}
\tilde{S} := \{\tilde{p}_{\theta}(x) = p_{\theta}(x)\lambda(\theta) : p_{\theta}\in S\},
\end{eqnarray}
where $\lambda$ is a probability distribution on $\Theta$. Then $\tilde{S}$ is a $k+1$-dimensional sub-manifold of $\tilde{\mathcal{P}}$. For $\tilde{p}_\theta, \tilde{p}_{\theta'}\in \tilde{\mathcal{P}}$, the Kullback-Leibler divergence (KL-divergence) between $\tilde{p}_\theta$ and $\tilde{p}_{\theta'}$ is given by
\begin{eqnarray}
	\label{eqn:KL-div}
	I(\tilde{p}_\theta\|\tilde{p}_{\theta'}) & = & \sum_x \tilde{p}_{\theta}(x) \log \frac{\tilde{p}_{\theta}(x)}{\tilde{p}_{\theta'}(x)} - \sum_x \tilde{p}_{\theta}(x) + \sum_x \tilde{p}_{\theta'}(x)\nonumber\\
	& = & \sum_x p_{\theta}(x)\lambda(\theta) \log \frac{p_{\theta}(x)\lambda(\theta)}{p_{\theta'}(x)\lambda(\theta')} - \lambda(\theta) + \lambda(\theta').\nonumber\\
\end{eqnarray}
We define a Riemannian metric $G^{(I)}(\theta) = [g_{i,j}^{(I)}(\theta)]$ on $\tilde{S}$ by
\begin{eqnarray}
	\nonumber
	\lefteqn{g_{i,j}^{(I)}(\theta)}\nonumber\\
	& := & - I[\partial_i\|\partial_j]\nonumber\\
	& = & \left. - \frac{\partial}{\partial \theta_i} \frac{\partial}{\partial \theta'_j} \sum_{x} p_{\theta}(x)\lambda(\theta) \log \frac{p_{\theta}(x)\lambda(\theta)}{p_{\theta'}(x)\lambda(\theta')} \right|_{\theta' = \theta}\nonumber\\
	& = & \sum_x \partial_i (p_{\theta}(x)\lambda(\theta))\cdot \partial_j \log (p _{\theta}(x)\lambda(\theta))\nonumber\\
	& = & \sum_x p_{\theta}(x)\lambda(\theta)\partial_i (\log p_{\theta}(x)\lambda(\theta))\cdot \partial_j (\log (p _{\theta}(x)\lambda(\theta)))\nonumber\\
	& = & \lambda(\theta)\sum_x p_{\theta}(x)[\partial_i (\log p_{\theta}(x)) + \partial_i (\log \lambda(\theta))]\nonumber\\
	& & \hspace*{2cm}\cdot [\partial_j (\log p_{\theta}(x)) + \partial_j (\log \lambda(\theta))]\nonumber\\
	& = & \lambda(\theta) \big\{E_{\theta}[\partial_i \log p_{\theta}(X)\cdot \partial_j \log p_{\theta}(X)]\nonumber\\
	& & \hspace*{2cm}\cdot  + \partial_i (\log \lambda(\theta))\cdot\partial_j (\log \lambda(\theta))\big\}\\
	\label{eqn:fisher-information-metric}
	& = & \lambda(\theta) \big\{g_{i,j}^{(e)}(\theta) + J_{i,j}^{\lambda}(\theta)\big\},
\end{eqnarray}
where
\begin{equation}
\label{eqn:Fisher_metric}
g_{i,j}^{(e)}(\theta) := E_{\theta}[\partial_i \log p_{\theta}(X)\cdot \partial_j \log p_{\theta}(X)],
\end{equation}
and
\begin{equation}
\label{eqn:J_matrix}
J_{i,j}^{\lambda}(\theta) := \partial_i (\log \lambda(\theta))\cdot\partial_j (\log \lambda(\theta)).
\end{equation} 
Let $G^{(e)}(\theta) := [g^{(e)}_{i,j}(\theta)]$ and $J^{\lambda}(\theta) := [J_{i,j}^{\lambda}(\theta)]$. Then 
\begin{equation}
\label{eqn:Bayesian_Fisher}
G^{(I)}(\theta) = \lambda(\theta)\big[G^{(e)}(\theta) + J^{\lambda}(\theta)\big].
\end{equation}
Notice that $G^{(e)}(\theta)$ is the usual {\em Fisher information matrix}. Also observe that $\tilde{\mathcal{P}}$ is a subset of $\mathbb{R}^{\tilde{\mathcal{X}}}$, where $\tilde{\mathcal{X}} := \mathcal{X}\cup\{a_{d+1}\}$. The tangent space at every point of $\tilde{\mathcal{P}}$ is $\mathcal{A}_0 := \{A\in\mathbb{R}^{\tilde{\mathcal{X}}} : \sum_{x\in\tilde{\mathcal{X}}}A(x) = 0\}$. That is, $T_p(\tilde{\mathcal{P}}) = \mathcal{A}_0$. We denote a tangent vector (that is, elements of $\mathcal{A}_0$) by $X^{(m)}$. The manifold $\tilde{\mathcal{P}}$ can be recognized by its homeomorphic image $\{\log \tilde{p}: \tilde{p}\in\tilde{\mathcal{P}}\}$ under the mapping $\tilde{p}\mapsto \log \tilde{p}$. Under this mapping the tangent vector $X\in T_{\tilde{p}}(\tilde{\mathcal{P}})$ can be represented $X^{(e)}$ which is defined by $X^{(e)}(x) = X^{(m)}(x)/\tilde{p}(x)$ and we define
\begin{equation*}
	T_{\tilde{p}}^{(e)}(\tilde{\mathcal{P}}) = \{X^{(e)} : X\in T_{\tilde{p}}(\tilde{\mathcal{P}})\} = \{A\in\mathbb{R}^{\tilde{\mathcal{X}}} : \mathbb{E}_{\tilde{p}}[A] = 0\}.
\end{equation*}

For the natural basis $\partial_i$ of a coordinate system $\theta = (\theta_i)$, $(\partial_i)^{(m)}_{\xi} = \partial_i \tilde{p}_\theta$ and $(\partial_i)^{(e)}_{\xi} = \partial_i \log \tilde{p}_\theta$.

With these notations, for any two tangent vectors $X,Y\in T_{\tilde{p}}(\tilde{\mathcal{P}})$, the Fisher metric in (\ref{eqn:fisher-information-metric}) can be written as
\begin{equation*}
	\langle X,Y\rangle_{\tilde{p}}^{(e)} = \mathbb{E}_{\tilde{p}}[X^{(e)}Y^{(e)}].
\end{equation*}

Let $\tilde{S}$ be a sub-manifold of $\tilde{\mathcal{P}}$ of the form as in (\ref{eqn:denormalized_manifold}), together with the metric $G^{(I)}$ as in (\ref{eqn:fisher-information-metric}). Let $T_{\tilde{p}}^*(\tilde{S})$ be the dual space (cotangent space) of the tangent space $T_{\tilde{p}}(\tilde{S})$ and let us consider for each $Y\in T_{\tilde{p}}(\tilde{S})$, the element $\omega_Y\in T_{\tilde{p}}^*(\tilde{S})$ which maps $X$ to $\langle X,Y\rangle^{(e)}$.  The correspondence $Y\mapsto \omega_Y$ is a linear map between $T_{\tilde{p}}(\tilde{S})$ and $T_{\tilde{p}}^*(\tilde{S})$. An inner product and a norm on $T_{\tilde{p}}^*(\tilde{S})$ are naturally inherited from $T_{\tilde{p}}(\tilde{S})$ by
\[
\langle \omega_X,\omega_Y\rangle_{\tilde{p}} := \langle X,Y\rangle^{(e)}_{\tilde{p}}
\]
and
\[
\|\omega_X\|_{\tilde{p}} := \|X\|_{\tilde{p}}^{(e)} = \sqrt{\langle X,X\rangle^{(e)}_{\tilde{p}}}.
\]
Now, for a (smooth) real function $f$ on  $\tilde{S}$, the \emph{differential} of $f$ at $\tilde{p}$, $(\text{d}f)_{\tilde{p}}$, is a member of $T_{\tilde{p}}^*(S)$ which maps $X$ to $X(f)$. The \emph{gradient} of $f$ at $\tilde{p}$ is the tangent vector corresponding to $(\text{d}f)_{\tilde{p}}$, hence satisfies
\begin{eqnarray}
	\label{eqn:differential_of_function}
	(\text{d}f)_{\tilde{p}}(X) = X(f) = \langle (\text{grad} f)_{\tilde{p}},X\rangle_{\tilde{p}}^{(e)},
\end{eqnarray}
and
\begin{eqnarray}
	\label{eqn:norm_of_differential}
	\|(\text{d}f)_{\tilde{p}}\|_{\tilde{p}}^2 = \langle (\text{grad}f)_{\tilde{p}},(\text{grad}f)_{\tilde{p}}\rangle_{\tilde{p}}^{(e)}.
\end{eqnarray}
Since $\text{grad}f$ is a tangent vector, we can write
\begin{equation}
	\label{eqn:grad-f}
	\text{grad}f = \sum\limits_{i=1}^k h_i \partial_i
\end{equation}
for some scalars $h_i$. Applying (\ref{eqn:differential_of_function}) with $X = \partial_j$, for each $j=1,\dots,k$, and using (\ref{eqn:grad-f}), we get
\begin{eqnarray*}
	(\partial_j)(f)
	& = & \left\langle \sum\limits_{i=1}^k h_i \partial_i, \partial_j\right\rangle^{(e)}\\
	& = & \sum\limits_{i=1}^k h_i \langle \partial_i, \partial_j\rangle^{(e)}\\
	& = & \sum\limits_{i=1}^k h_i g_{i,j}^{(e)}, \quad j = 1, \dots, k.
\end{eqnarray*}
From this, we have
\[
[h_1,\dots,h_k]^T = \left[G^{(e)}\right]^{-1}[\partial_1(f),\dots,\partial_k(f)]^T,
\]
and so
\begin{equation}
	\label{eqn:grad-coeff-equation}
	\text{grad}f = \sum\limits_{i,j} (g^{i,j})^{(e)}\partial_j(f) \partial_i.
\end{equation}
From (\ref{eqn:differential_of_function}), (\ref{eqn:norm_of_differential}), and (\ref{eqn:grad-coeff-equation}), we get
\begin{eqnarray}
	\label{differential_and_metric}
	\|(\text{d}f)_{\tilde{p}}\|_{\tilde{p}}^2 = \sum\limits_{i,j} (g^{i,j})^{(e)}\partial_j(f) \partial_i(f)
\end{eqnarray}
where $(g^{i,j})^{(I)}$ is the $(i,j)$th entry of the inverse of $G^{(I)}$.

The above and the following results are indeed an extension of \cite[Sec.2.5.]{amari2000methods} to $\tilde{\mathcal{P}}$.
\begin{thm}[\cite{amari2000methods}]
	Let $A:\mathbb{X}\to\mathbb{R}$ be any mapping (that is, a vector in $\mathbb{R}^{\mathbb{X}}$. Let $E[A]:\mathcal{\tilde{P}}\to \mathbb{R}$ be the mapping $\tilde{p}\mapsto E_{\tilde{p}}[A]$. We then have
	\begin{eqnarray}
		\label{eqn:variance_eq_norm_of_differential}
		\text{Var}(A) =  \|(\text{d}E_{\tilde{p}}[A])_{\tilde{p}}\|_{\tilde{p}}^2.
	\end{eqnarray}
	$\hfill \IEEEQEDopen$
\end{thm}
\begin{IEEEproof}
	For any tangent vector $X\in T_{\tilde{p}}(\mathcal{\tilde{P}})$,
	\begin{eqnarray}
		\label{eqn:tangent_acting_on_expectation}
		X(E_{\tilde{p}}[A])
		& = & \sum\limits_x X(x)A(x)\nonumber\\
		& = & E_{\tilde{p}}[X_{\tilde{p}}^{(e)} \cdot A]\\
		& = & E_{\tilde{p}}[X_{\tilde{p}}^{(e)}(A-E_{\tilde{p}}[A])].
	\end{eqnarray}
	Since $A-E_{\tilde{p}}[A]\in T_{\tilde{p}}^{(e)}(\mathcal{\tilde{P}})$, there exists $Y\in T_{\tilde{p}}(\mathcal{\tilde{P}})$ such that $A-E_{\tilde{p}}[A] = Y_{\tilde{p}}^{(e)}$, and $\text{grad}(E[A]) = Y$. Hence we see that
	\begin{eqnarray*}
		\lefteqn{\|(\text{d}E[A])_{\tilde{p}}\|_{\tilde{p}}^2}\\
		& & \hspace{-0.4cm} = E_p[Y_{\tilde{p}}^{(e)}Y_{\tilde{p}}^{(e)}]\\
		& & \hspace{-0.4cm} = E_{\tilde{p}}[(A-E_{\tilde{p}}[A])^2].
	\end{eqnarray*}
\end{IEEEproof}

\begin{corollary}[\cite{amari2000methods}]
	\label{cor:c_r_inequality}
	If $S$ is a submanifold of $\mathcal{\tilde{P}}$, then
	\begin{eqnarray}
		\label{eqn:variance_ge_norm_of_differential}
		\text{Var}_{\tilde{p}}[A] \ge \|(\text{d}E[A]|_{S})_{\tilde{p}}\|_{\tilde{p}}^2
	\end{eqnarray}
	with equality iff $$A-E_{\tilde{p}}[A]\in \{X_{\tilde{p}}^{(e)} : X\in T_{\tilde{p}}(S)\} =: T_{\tilde{p}}^{(e)}(S).$$ $\hfill \IEEEQEDopen$
\end{corollary}
\begin{IEEEproof}
	Since $(\text{grad }E[A]|_{S})_{\tilde{p}}$ is the orthogonal projection of $(\text{grad }E[A])_{\tilde{p}}$ onto $T_{\tilde{p}}(S)$, the result follows from the theorem.
\end{IEEEproof}

\section{Derivation of Error Bounds}
\label{sec:error_bounds}
We state our main result in the following theorem.

\begin{theorem}
\label{thm:main}
	Let $S$ and $\tilde{S}$ be as in (\ref{eqn:denormalized_manifold}). Let $\widehat{\theta}$ be an estimator of $\theta$. Then
	
	\begin{enumerate}
		\item[(a)] {\em Bayesian Cram\'{e}r-Rao}: 
		\begin{equation}
		\label{eqn:Bayesian_Cramer-Rao}
		\mathbb{E}_\lambda\big[\text{Var}_\theta(\widehat{\theta})\big] \ge \big\{\mathbb{E}_\lambda [G^{(I)}(\theta)]\big\}^{-1}, 
		\end{equation}
		where $\text{Var}_\theta(\widehat{\theta}) = [\text{Cov}_{\theta}(\widehat{\theta}_i(X),\widehat{\theta}_j(X))]$  is the covariance matrix and $G^{(I)}(\theta)$ is as in (\ref{eqn:Bayesian_Fisher}). (In (\ref{eqn:Bayesian_Cramer-Rao}), we use the usual convention that, for two matrices $A$ and $B$, $A\ge B$ means that $A-B$ is positive semi-definite.)
		 
		\item[(b)] {\em Deterministic Cram\'{e}r-Rao} (unbiased): If $\widehat{\theta}$ is an unbiased estimator of $\theta$, then
		\begin{equation}
		\label{eqn:Deter_Cramer_Rao}
		\text{Var}_\theta[\widehat{\theta}] \ge [G^{(e)}(\theta)]^{-1}.
		\end{equation}
		
		\item[(c)] {\em Deterministic Cram\'{e}r-Rao (biased)}: For any estimator $\widehat{\theta}$ of $\theta$,
		\begin{eqnarray*}
		\text{MSE}_\theta[\widehat{\theta}] \ge (\textbf{1} + B'(\theta)) [G^{(e)}(\theta)]^{-1}(\textbf{1} + B'(\theta))\\
		+ b(\theta)b(\theta)^T,
		\end{eqnarray*}
		where $b(\theta) = (b_1(\theta),\dots,b_k(\theta))^T := \mathbb{E}_\theta[\widehat{\theta}]$ is the bias and $\textbf{1} + B'(\theta)$ is the matrix whose $(i,j)$th entry is $0$ if $i\neq j$ and is $(1+\partial_i b_i(\theta))$ if $i=j$.
		
		\item[(d)] {\em Barankin Bound}: (Scalar case) If $\widehat{\theta}$ be an unbiased estimator of $\theta$, then
		\begin{equation}
		\label{eqn:Barankin_Bound}
		\text{Var}_\theta[\widehat{\theta}] \ge \sup_{n,a_l,\theta^{(l)}}\frac{\Big[\sum\limits_{l=1}^n a_l(\theta^{(l)}-\theta)\Big]^2}{\sum\limits_x \Big[\sum\limits_{l=1}^n a_l L_{\theta^{(l)}}(x)\Big]^2 p_\theta(x)},
		\end{equation}
		where $L_{\theta^{(l)}}(x) := {p_{\theta^{(l)}}(x)}{p_\theta(x)}$ and the supremum is over all $a_1,\dots,a_n\in\mathbb{R}$, $n\in\mathbb{N}$, and $\theta^{(1)},\dots,\theta^{(n)}\in\Theta$.
	\end{enumerate}
\end{theorem}

\begin{IEEEproof}
	\begin{enumerate}
		\item[(a)] Let $A = \sum_{i=1}^{k}c_i\widehat{\theta}_i$, where $\widehat{\theta} = (\widehat{\theta}_1,\dots,\widehat{\theta}_k)$ is an unbiased estimator of $\theta$, in corollary \ref{cor:c_r_inequality}. Then, from (\ref{eqn:variance_ge_norm_of_differential}), we have
		\begin{equation*}
		\sum\limits_{i,j} c_i c_j \text{Cov}_{\tilde{\theta}}(\widehat{\theta}_i,\widehat{\theta}_j)\ge \sum\limits_{i,j} c_i c_j (g^{(I)})^{i,j}(\theta).
		\end{equation*}
		This implies that
			\begin{equation}
			\label{eqn:cramer_rao_general}
		\lambda(\theta)\sum\limits_{i,j} c_i c_j \text{Cov}_{\theta}(\widehat{\theta}_i,\widehat{\theta}_j)\ge \sum\limits_{i,j} c_i c_j (g^{(I)})^{i,j}(\theta).
		\end{equation}
		Hence, taking expectation on both sides with respect to $\lambda$, we get
		\begin{equation*}
		\sum\limits_{i,j} c_i c_j \mathbb{E}_\lambda\big[\text{Cov}_{\theta}(\widehat{\theta}_i,\widehat{\theta}_j)\big]\ge \sum\limits_{i,j} c_i c_j \mathbb{E}_\lambda\big[(g^{(I)})^{i,j}(\theta)\big].
		\end{equation*}
		That is,
		\begin{equation*}
		\mathbb{E}_\lambda\big[\text{Var}_\theta(\widehat{\theta})\big] \ge \mathbb{E}_\lambda [G^{(I)}(\theta)^{-1}].
		\end{equation*}
		But $\mathbb{E}_\lambda [G^{(I)}(\theta)^{-1}] \ge \big\{\mathbb{E}_\lambda [G^{(I)}(\theta)]\big\}^{-1}$ by \cite{GrovesRothenberg1969Biometrika}. This proves the result.
		
		\item[(b)] This follows from (\ref{eqn:cramer_rao_general}) by taking $\lambda(\theta) = 1$.
		
        \item[(c)] Let us first observe that $\widehat{\theta}$ is an unbiased estimator of $\theta + b(\theta)$. Let $A = \sum_{i=1}^{k} c_i\widehat{\theta}_i$ as before. Then $\mathbb{E}[A] = \sum_{i=1}^{k}c_i(1+b_i(\theta))$. Then, from corollary \ref{cor:c_r_inequality} and (\ref{eqn:variance_ge_norm_of_differential}), we have
		\begin{equation*}
		 \text{Var}_\theta[\widehat{\theta}] \ge (\textbf{1} + B'(\theta)) [G^{(e)}(\theta)]^{-1}(\textbf{1} + B'(\theta))
		\end{equation*}
		But $\text{MSE}_\theta[\widehat{\theta}] = \text{Var}_\theta[\widehat{\theta}] + b(\theta)b(\theta)^T$. This proves the assertion.
        
		\item[(d)] For fixed $a_1,\dots,a_n\in\mathbb{R}$ and $\theta^{(1)},\dots,\theta^{(n)}\in\Theta$, let us define a metric by the following formula
		\begin{equation}
		\label{Eqn:Metric_Barankin}
		g(\theta) := \sum\limits_x \Big[\sum\limits_{l=1}^n a_l L_{\theta^{(l)}}(x)\Big]^2 p_\theta(x).
		\end{equation}
		Let $f$ be the mapping $p\mapsto\mathbb{E}_p[A]$. Let $A(\cdot) = \widehat{\theta}(\cdot) - \theta$, where $\widehat{\theta}$ is an unbiased estimator of $\theta$. Then the partial derivatives of $f$ in (\ref{differential_and_metric}) equals to
		\begin{eqnarray*}
		\sum\limits_x(\widehat{\theta}(x) - \theta)\Big(\sum\limits_{l=1}^na_lL_{\theta^{(l)}}(x)\Big)p_\theta(x)\\
		&&\hspace*{-4cm} = \sum\limits_{l=1}^n a_l \Big(\sum\limits_x(\widehat{\theta}(x) - \theta)\frac{p_{\theta^{(l)}}(x)}{p_\theta(x)}p_\theta(x)\Big)\\
		&&\hspace*{-4cm} = \sum\limits_{l=1}^n a_l (\theta^{(l)} - \theta).
		\end{eqnarray*}
		Hence from (\ref{differential_and_metric}) and corollary \ref{cor:c_r_inequality}, we have
		 \begin{equation*}
		 \text{Var}_\theta[\widehat{\theta}] \ge \frac{\Big[\sum\limits_{l=1}^n a_l(\theta^{(l)}-\theta)\Big]^2}{\sum\limits_x \Big[\sum\limits_{l=1}^n a_l L_{\theta^{(l)}}(x)\Big]^2 p_\theta(x)}
		 \end{equation*}
		 Since $a_l$ and $\theta^{(l)}$ are arbitrary, taking supremum over all $a_1,\dots,a_n\in\mathbb{R}$, $n\in\mathbb{N}$, and $\theta^{(1)},\dots,\theta^{(n)}\in\Theta$, we get (\ref{eqn:Barankin_Bound}).
	\end{enumerate}
\end{IEEEproof}	

\section{Conclusion}
\label{sec:conc}
We have shown that our Theorem \ref{thm:main} provides a general information geometric characterization of the statistical manifolds linking them to the Bayesian CRLB for vector parameters; the extension to estimators of measurable functions of the parameter $\theta$ is trivial. We exploited the general definition of Kullback-Leibler divergence when the probability densities are not normalized. This is an improvement over Amari’s work \cite{amari2000methods} on information geometry which only dealt with the notion of deterministic CRLB of scalar parameters. Further, we proposed an approach to arrive at the Barankin bound thereby shedding light on the relation between the threshold effect and information geometry. Both of our improvements enable usage of information geometric approaches in critical scenarios of biased estimators and low SNRs. This is especially useful in the analyses of many practical problems such as radar and communication. In future investigations, we intend to explore these methods further especially in the context of the threshold effect.

%% References:
\balance
\bibliographystyle{IEEEtran}
\bibliography{main}

\end{document}